\documentclass[aps,pre,groupedaddress,showpacs,twocolumn,floatfix]{revtex4}
\usepackage{graphicx}
\usepackage{dcolumn}
\usepackage{bm}
\usepackage{amssymb}
\usepackage{amsmath}
\usepackage{subfigure}
\def\LM#1#2{\left|\begin{array}{l}{#1}\\[1ex]{#2}\end{array}\right.}
\begin{document}

\title{Reaction Front in an $A+B\to C$ Reaction-Subdiffusion Process}
\author{S. B. Yuste$^{\dag}$, L. Acedo$^{\dag}$, and Katja
Lindenberg$^{\ddag}$ }
\affiliation{
$^{\dag}$Departamento de F\'{\i}sica, Universidad de Extremadura, 
E-06071 Badajoz, Spain\\
$^{\ddag}$Department of Chemistry and Biochemistry, and Institute for
Nonlinear Science, University of California San Diego, 9500 Gilman Dr.,
La Jolla, CA 92093-0340, USA
}

\begin{abstract}
We study the reaction front for the process $A+B\to C$ 
in which the reagents move subdiffusively.  Our theoretical description
is based on a fractional reaction-subdiffusion equation in which both 
the motion and the reaction terms are affected by the subdiffusive
character of the process.  We design numerical simulations to check our
theoretical results, describing the simulations in some detail because
the rules necessarily differ in important respects from those used in
diffusive processes.  Comparisons between theory and simulations are on
the whole favorable, with the most difficult quantities to capture being
those that involve very small numbers of particles.  In particular, we
analyze the total number of product particles, the width of the
depletion zone, the production profile of product and its width, as well as
the reactant concentrations at the center of the 
reaction zone, all as a function of time.  We also analyze
the shape of the product profile as a function
of time, in particular its unusual behavior at the center of the
reaction zone.
\end{abstract}

\pacs{82.33.-z, 82.40.-g, 02.50.Ey, 89.75.Da}

\maketitle

\section{Introduction}
\label{sec:1}
It is very well known that
diffusion-limited binary reactions in low dimensions may
lead to the spontaneous appearance of spatial order and spatial
structures,
and to associated ``anomalous'' rate laws for the global densities
of the reacting species.  For example, the reactions
$A + A \rightarrow
C$ (some selected references of many in the literature
are~\cite{book,TorMcCJPC,Spouge,oldA+A,ours,KrebsEtAl,MasAvrPLA,wenew,ManAvr,HenHin,AbaFriNic})
and $A+B \rightarrow C$ (again, some of many references
are~\cite{book,ovchinnikov,toussaint,kang,science,kanno1,clement1,argyrakis2,we,bramson,leyvraz1,zumofen,we2}) under ``normal'' circumstances are
described by
second-order rate laws, whereas the asymptotic rate law for the former
reaction is of apparent order $(1 + 2/d)$ for dimensions $d < 2$,
and for the mixed reaction it is of apparent order $(1+4/d)$ for $d<4$.
The slow-down implied by the
higher order is a consequence of the rapid deviation of the
spatial distribution of reactants from a random distribution.
This is in turn a consequence of the fact that diffusion is not an
effective mixing mechanism in low dimensions. 

To design an experiment in a constrained geometry in order to measure
these anomalies is not at all simple, especially for the mixed
reaction~\cite{kopelman,gelfree,baroud}.
It is simpler for the $A+A$ problem because a number of appropriate
non-chemical species can be identified that essentially undergo the
simplest annihilation reaction or variants thereof.  Examples include
exciton annihilation experiments in one-dimensional pores and in effectively
one-dimensional polymer wires~\cite{science}, excited molecule
naphthalene fusion and quenching experiments in one-dimensional
pores~\cite{experiment},
and kink-antikink simulations in one dimension~\cite{habib}. 
Experimental observations of the $A+B$ anomalies instead
generally involve \emph{reaction fronts}. 
Early on G\'alfi and R\'acz~\cite{galfi} and later
others~\cite{jiang,cornell1,cornell2,redner,larralde,droz,cardy,krapivsky,kozak}
recognized that the kinetic
anomalies in the homogeneous systems would be reflected in the
evolution of reaction fronts. On the basis of scaling arguments,
later made more rigorous, a number of exponents were deduced to
characterize this evolution.
The first experiments confirming these results were carried
out with species $A$ and $B$ diffusing in a gel contained in a thin
capillary, with one species initially occupying one side, the other
occupying the other side, and a sharp front between them~\cite{kopelman}. 
More recent experiments have been carried out in gel-free
systems~\cite{gelfree,baroud}.  The evolution of the reactant fronts and of
the product of the reaction in these experiments both reflect the
kinetic anomalies.  

In this paper we extend the front analysis to subdiffusive reactions.
Subdiffusive motion is characterized by a mean square displacement that
varies sublinearly with time,
\begin{equation}
\langle r^2(t)\rangle \sim \frac{2K_\gamma}{\Gamma(1+\gamma)} t^\gamma ,
\label{meansquaredispl}
\end{equation}
with $0<\gamma<1$.  For ordinary diffusion $\gamma=1$, and $K_1\equiv D$
is the ordinary diffusion coefficient.
We argue for the importance of this generalization on a number of
grounds.  First, there exists a huge literature on systems that deviate
from diffusive behavior and are instead characterized 
by motion all the way from subdiffusive to
superdiffusive~\cite{several1,several2}.
Subdiffusive motion is particularly important in the context of complex
systems such as glassy and disordered materials, in which pathways
are constrained for geometric or for energetic reasons.  It is also
particularly germane
to the way in which experiments in low dimensions have to
be carried out.  Such experiments must avoid any active or
convective or advective mixing so as to ensure that any
mixing is only a consequence of diffusion.  To accomplish this
usually requires the use of gel substrates and/or highly constrained
geometries (the first gel-free experiments were carried out
recently~\cite{gelfree,baroud}).  Under
these circumstances it is not clear whether the motion of the species is
actually diffusive, or if it is in fact \emph{subdiffusive}.  Indeed, a
recent detailed discussion on ways to extract accurate parameters
and exponents from such experiments concludes that at least
the experiments presented in that work, carried out in a gel,
reflect subdiffusive rather than diffusive motion~\cite{recent}. 

We recently solved the $A+A$ reaction-subdiffusion problem in one
dimension~\cite{yuste}. To solve this problem, we generalized
methods first applied to the reaction-diffusion $A+A$ problem.
For the $A+A\to A$ problem the method of intervals allows 
an exact formulation in terms of intervals on the line that are empty
of $A$
particles~\cite{oldA+A,MasAvrPLA,ManAvr,KrebsEtAl,HenHin,AbaFriNic}. 
The distribution of intervals evolves linearly, and
therefore one can find an exact solution.  In the reaction-diffusion
problem the description involves a diffusion equation, while the
reaction-subdiffusion problem involves a subdiffusion
equation~\cite{yuste}; both can
be solved exactly.  For the $A+A\to C$ problem one uses instead the
odd/even parity method~\cite{MasAvrPLA,wenew}, whereby one keeps
track of the parity of the
number of particles in an interval.  The associated distribution again
satisfies a linear diffusion or subdiffusion~\cite{yuste} equation.  

Before these exact methods were developed, it was customary to model
these and other binary reactions by writing down a reaction-diffusion
equation for each species in the reaction.  Such equations typically
contain a diffusion term and a binary reaction term, the latter being
some truncated form of a two-particle distribution function.  For
instance, in the $A+B$ reaction the typical reaction term simply
involves the product of the local concentrations of reactants,
$-ka({\bf r},t)b({\bf r},t)$~\cite{book}.  In the $A+A$ reaction one has
to be slightly
more careful because in writing, for example, $-ka^2(\bf r,t)$ one must
be careful not to include spurious self-reaction
contributions~\cite{ours,wenew}.  Once the $A+A$ exact models were
developed that did not require one to explicitly write a reaction term,
it was possible to analyze the accuracy of the approximate
truncations~\cite{lin}. Also, it was not necessary to consider the
generalization of the reaction term to the subdiffusive case since the
exact methods could be generalized directly.

The situation is more complicated for the $A+B$ problem, because no such
exact formulations or solutions have been developed in this case.  There
is a large literature on the reaction-diffusion problem with different
truncation schemes to represent the reaction term, but
the literature on the reaction-subdiffusion problem is far more recent and
relatively unsettled.  In particular, at the current stage of
development of this
problem it \emph{is} necessary to think about how to (approximately)
model the reaction term.

In Sec.~\ref{sec:2} we present a discussion of the model to be used for
the description of the $A+B$ reaction-subdiffusion problem.
Having arrived at a particular set of fractional equations.
We apply a scaling theory to these equations akin to that
of G\'alfi and R\'acz~\cite{galfi}, but now for a subdiffusive front. 
To support
the theoretical conclusions, it is necessary to perform numerical
simulations, which is not a trivial matter for a problem involving
subdiffusion.  In
Sec.~\ref{sec3}
we discuss our Monte Carlo simulation methods.  Section~\ref{sec4} is
a compendium and comparison of numerical and theoretical results.  Some
closing comments are presented in Sec.~\ref{sec5}.

\section{The Model}
\label{sec:2}
We start with a system of $A$ particles on one side and $B$ particles on
the other of a sharp linear front, defined to lie 
perpendicular to the $x$ axis.  The particles diffuse and react
with a given probability upon encounter.  A standard mean-field model for the
evolution of the concentrations $a(x, t)$ and $b(x, t)$
of $A$ and $B$ particles along $x$ is given by the reaction-diffusion
equations
\begin{equation}
\label{difmft}
\begin{array}{rcl}
\displaystyle\frac{\partial}{\partial t} a(x,t)&=& D
\displaystyle\frac{\partial^2}{\partial x^2} a(x,t)-k a(x,t) b(x,t) \\
\noalign{\smallskip}
\displaystyle\frac{\partial}{\partial t} b(x,t)&=& D
\displaystyle\frac{\partial^2}{\partial x^2} b(x,t)-k a(x,t) b(x,t) \; ,
\end{array}
\end{equation}
where $D$ is the diffusion coefficient assumed to be equal for the two
species.  The initial conditions are that $a(x,t)=const = a_0$ for $x<0$ and 
$a(x,t)=0$ for $x\geq0$.  Similarly, $b(x,t)=const = b_0$ for $x>0$ and
$b(x,t)=0$ for $x\leq0$. With these conditions, no matter the
dimensionality of the system, the system of equations is effectively
one-dimensional.
The front problem was first analyzed via a scaling
description~\cite{galfi} and later refined by a large number of
authors using more rigorous theoretical and careful numerical
approaches~\cite{jiang,cornell1,cornell2,redner,larralde,droz,cardy,krapivsky,kozak}.
One upshot of the extensive work is that $d=2$ is a critical dimension
for the mean field description to be appropriate. Below $d=2$
one must take into account fluctuations, neglected in this description,
that completely change the outcome of the analysis. 
A particularly transparent argument for this critical dimension was
provided by Krapivsky~\cite{krapivsky}.  He argued that the reaction
constant in the mean-field reaction rate $r=kab$ should in general
depend on the diffusion constant $D$ and the radius $R$ of the reacting
particles.  Dimensional analysis gives $k\sim D R^{d-2}$, but on physical
grounds one expects the reaction rate constant to be an increasing
function of the radius $R$.  The conclusion is that the mean field model
can therefore not be valid for $d<2$.  While it has been assumed that
the mean field model holds for the critical dimensions $d=2$, Krapivsky
finds logarithmic corrections that have also been observed in
simulations~\cite{cornell1}.
In our analysis and
simulations we will take $d=2$ (which turns out to be the critical
dimension for the subdiffusive problem as well) and will therefore not
deal with the lower-dimensional fluctuation effects.  
In this first study we will not deal with logarithmic corrections.

In order to generalize the reaction-diffusion problem to
reaction-subdiffusion, we must deal with the subdiffusive 
motion of the particles (generalization of the first term in
Eq.~(\ref{difmft})) and with their reaction rate law (second term).
We discuss each separately.

Subdiffusion is not modeled in a universal way in the literature. 
Among the more successful approaches to the subdiffusion problem
have been continuous time random walks with non-Poissonian
waiting time distributions~\cite{Kehr,Barkai,Sokolov}, and fractional
dynamics approaches in which the diffusion operator is replaced by
a generalized \emph{fractional} diffusion 
operator~\cite{several1,Barkai,severalmore1,severalmore2,Sch}.
The relation between the two has also been
discussed~\cite{several1,Barkai,severalmore2}.
In particular, the fractional dynamics formulation that leads to the
mean square displacement~(\ref{meansquaredispl}) can be associated with a
continuous time random walk with a waiting time distribution between
steps which at long times behaves as
\begin{equation}
\psi(t)\sim t^{-\gamma-1}.
\label{waitingtime}
\end{equation}
We adopt the fractional dynamics approach,
and comment later on some issues associated with it that must carefully
be considered in the context of numerical simulations.  We thus replace
Eq.~(\ref{difmft}) with the set of reaction-subdiffusion equations
\begin{equation}
\label{subdifmft}
\begin{array}{rcl}
\displaystyle\frac{\partial}{\partial t} a(x,t)&=&
K_\gamma
\displaystyle ~_{0}D_{t}^{1-\gamma }
\frac{\partial^2}{\partial x^2} a(x,t)-R_\gamma(x,t) \\
\noalign{\smallskip}
\displaystyle\frac{\partial}{\partial t} b(x,t)&=& 
K_\gamma
\displaystyle ~_{0}D_{t}^{1-\gamma }
\frac{\partial^2}{\partial x^2} b(x,t)-R_\gamma(x,t) \; ,
\end{array}
\end{equation}
where $K_\gamma$ is the generalized diffusion coefficient
that appears in Eq.~(\ref{meansquaredispl}), and 
$~_{0}D_{t}^{1-\gamma } $ is the Riemann-Liouville operator,
\begin{equation}
~_{0}D_{t}^{1-\gamma } f(x,t)=\frac{1}{\Gamma(\gamma)}
\frac{\partial}{\partial t} \int_0^t d\tau
\frac{f(x,\tau)}{(t-\tau)^{1-\gamma}}.
\label{LR}
\end{equation}
The reaction term $R_\gamma(x,t)\equiv R$ will be discussed subsequently,
because certain aspects of the problem are independent of the specific
form of this term.    

\subsection{Scaling independent of reaction term}
\label{model1}
As the reaction proceeds, a depletion zone develops around the front. 
This is the region where the concentrations of reactants are
significantly smaller than
their initial values. How the width $W_d$ evolves with time is one of
the measures typically used to characterize the process.  Within this
depletion zone lies the so-called reaction zone, the region where the
concentration $c(x,t)$ of the product $C$ is
appreciable.  This concentration profile has a width $w$ whose variation
with time is another characteristic of the evolving reaction. The
evolution of the production rate of $C$ (which determines the height
of the profile of $c(x,t)$ in the reaction zone) is a third measure of
the process.  To find these time dependences
we adapt the original scaling approach~\cite{galfi,cornell1}
to the subdiffusive case, and assume the scaling forms
\begin{equation}
\label{abscaling}
\begin{array}{rcl}
a(x,t)&=& t^{-\theta} \hat{a} (x t^{-\alpha}) \\
\noalign{\smallskip}
b(x,t)&=& t^{-\theta} \hat{b} (x t^{-\alpha})
\end{array}
\end{equation}
for the concentrations and
\begin{equation}
\label{Rgamma}
R_\gamma(x,t)=t^{-\mu} \hat{R}_\gamma (x t^{-\alpha}) 
\end{equation}
for the reaction term.  The exponents $\theta$, $\alpha$, and $\mu$
are to be determined from three relations.  The scaling forms
are only valid for $x\ll W_d$,
that is, well within the depletion zone.  

Two of the three relations needed to fix the scaling exponents do not
require further specification of the reaction term.  
Since the reaction zone increases more slowly than the width of the
depletion zone (an assumption that ex post turns out to be correct), we
can focus on the concentration difference $u(x,t)=a(x,t)-b(x,t)$ to
deduce the width of the latter.  The reaction term drops out when one
subtracts the equations in Eq.~(\ref{subdifmft}), and its form therefore
does not matter at this point.  Generalizing the procedure
in~\cite{galfi}, one can scale
the resulting equation by measuring concentrations in units of $a_0$,
time in units of $\tau=1/(ka_0)$, and length in units of
$l=(K_\gamma\tau^\gamma)^{1/2}$, so that the equation is simply
\begin{equation}
\displaystyle\frac{\partial}{\partial t} u(x,t)=
\displaystyle ~_{0}D_{t}^{1-\gamma }
\frac{\partial^2}{\partial x^2} u(x,t)
\end{equation}
and the only control parameter is $q=b_0/a_0$ in the initial condition:
\begin{equation}
\label{difference}
\begin{array}{rcll}
u(x,0)&=& 1&{\rm for}~~~ x<0 \\
\noalign{\smallskip}
u(x,0)&=& -q &{\rm for}~~~ x>0.
\end{array}
\end{equation}
The solution is 
\begin{equation}
u(x,t)= -q +\frac{1+q}{2}
H^{1,0}_{1,1}\left[\frac{x}{t^{\gamma/2} }
        \LM{(1,\frac{\gamma}{2})}{(0,1)}   \right].
\label{ctPeriodica}
\end{equation}
where $H^{1,0}_{1,1}$ is the Fox
H-function~\cite{Sch,MathaiSaxena,AbraMPL}.
When $\gamma=1$ this reduces to the diffusion result~\cite{galfi}
\begin{equation}
u(x,t)= -q +\frac{1+q}{2}\;{\rm erfc}\left( \frac{x}{2t^{1/2}}\right).
\end{equation}
From Eq.~(\ref{ctPeriodica}) we see that the width of the depletion
zone scales as
\begin{equation}
W_d\sim t^{\gamma/2},
\label{width}
\end{equation}
i.e., $\partial a(x,t)/\partial x
\sim \partial b(x,t)/\partial x \sim t^{-\gamma/2}$.
Then, from Eq.~(\ref{abscaling}), the following relation between scaling
exponents follows immediately: 
\begin{equation}
\theta + \alpha = \frac{\gamma}{2}.
\label{scaling1}
\end{equation}

The second relation follows from the fact that the concentration
gradient of $A$ and $B$ leads to a flux of particles toward the reaction
region. The assumption that the reaction is fed by these particle currents
then leads to the quasistationary form in the reaction zone
\begin{equation}
\label{subdifmftstat}
\begin{array}{rcl}
0&=&
K_\gamma
\displaystyle ~_{0}D_{t}^{1-\gamma }
\frac{\partial^2}{\partial x^2} a(x,t)-R_\gamma(x,t) \\
\noalign{\smallskip}
0&=&
K_\gamma
\displaystyle ~_{0}D_{t}^{1-\gamma }
\frac{\partial^2}{\partial x^2} b(x,t)-R_\gamma(x,t) \; ,
\end{array}
\end{equation}
which requires that
\begin{equation}
\mu=\theta+2\alpha+1-\gamma\;.
\label{scaling2}
\end{equation}

For the width of the reaction zone to grow more slowly than the
depletion zone caused by subdiffusion requires that
\begin{equation}
\alpha < \gamma/2\;.
\label{ineq}
\end{equation}
On the other hand, the quasistationarity condition requires that 
\begin{eqnarray}
K_\gamma
\displaystyle ~_{0}D_{t}^{1-\gamma }
\frac{\partial^2}{\partial x^2} a(x,t) &\sim&
t^{-(\theta+2\alpha+1-\gamma)} \nonumber\\
&\gg& 
\displaystyle\frac{\partial}{\partial t} a(x,t) \sim t^{-(\theta+1)},
\end{eqnarray}
which again leads to Eq.~(\ref{ineq}).

Equations~(\ref{scaling1}) and (\ref{scaling2}) combined lead to the
relation $\alpha-\mu=\gamma/2-1$
that is easily checked by numerical simulations, since it is
determined by
the production rate of $C$.  The rate of change of the total amount of
product, $dN_C/dt$, is given by the integral of the reaction rate over
the reaction zone,
\begin{eqnarray} 
\frac{dN_C}{dt} = \int_{\substack{{\rm reaction} \\  {\rm zone}}}
dx\; R_\gamma(x,t)
&\sim& t^{-\mu}
\int_{\substack{{\rm reaction} \\  {\rm zone}}}
dx\; \hat{R}_\gamma(x/t^{\alpha}) \nonumber\\
&\sim& t^{-(\mu-\alpha)}
\sim t^{\gamma/2 -1},
\label{crate}
\end{eqnarray}
that is,
\begin{equation}
N_C(t)\sim t^{\gamma/2}.
\label{ctot}
\end{equation}
We stress that this total amount of product as a function of time, which is
numerically more robust than its derivative, 
is predicted to grow as $t^{\gamma/2}$ \emph{regardless} of the
specific form of the reaction term.

Another accessible quantity that is independent of $R_\gamma$ is the
location $x_f$ of the point at which the production rate of $C$ is largest.
This should occur where $a(x,t)\sim b(x,t)$, that is, $u(x_f,t)\sim
0$.  The time dependence of this equimolar point is found from
Eq.~(\ref{difference}) to be
\begin{equation}
x_f(t)=K_ft^{\gamma/2}
\end{equation}
where $K_f$ is determined from the equation
\begin{equation}
\frac{2q}{1+q}= 
H^{10}_{11}\left[K_f \LM{(1,\frac{\gamma}{2})}{(0,1)} \right].
\end{equation}

\subsection{Choice of reaction term and resultant scaling}
\label{model2}

Further relations involving the scaling exponents aimed at their
expression in terms of model quantities require specification of the
reaction term. There is a varied literature on this subject, based on a
number of different assumptions~\cite{wearne,sung,seki1,seki2}. 
Most do not associate a memory with the reaction term. Some
assume that, as in the case of ordinary diffusion, reactions can simply
be modeled by a space-dependent form of the law of mass action, e.g.,
by setting $R=ka(x,t)b(x,t)$.  Some of these assumptions may be
appropriate if the reaction is very rapid, but not if many encounters
between reactants are required for the reaction to occur. 

We adopt the viewpoint put forth in a recent theory developed for
geminate recombination~\cite{seki1,seki2} but,
as the authors themselves point out, much more broadly applicable.
This theory
goes back to the continuous time random walk picture from which the
fractional diffusion equation can be obtained, and considers
\emph{both} the motion and the reaction in this framework. 
In the context of geminate recombination the authors define a reaction
zone and argue that a geminate pair within the reaction zone will not
necessarily react for any finite intrinsic reaction rate
(which they call $\gamma_{rc}$) because one of the
particles may leave the zone before a reaction takes place. The dynamics
of leaving the reaction zone is ruled by the waiting time distribution
$\psi_{out}(t)=\psi(t)e^{-\gamma_{rc}t}$ where $\psi(t)$ is the waiting
time that regulates the rest of the dynamics [cf.
Eq.~(\ref{waitingtime})], and therefore the reaction
rate will acquire a memory that arises from the same source as the
memory associated with the subdiffusive motion.  In the continuum limit
this model then leads to a reaction-subdiffusion equation in which both
contributions have a memory.  Seki et al. obtain a subdiffusion-reaction
equation which at long times corresponds to choosing a reaction term of
the form
\begin{equation}
R_\gamma(x,t)=k ~_{0}D_{t}^{1-\gamma }a(x,t)b(x,t).
\label{specific}
\end{equation}
Here ``long times'' set in very quickly if the reaction zone is narrow
and the intrinsic reaction rate small.  As noted earlier, although the
derivation is specifically for geminate recombination, the arguments can
be generalized. 

Our full reaction-subdiffusion starting equations
on which the remainder of this paper is based then are 
\begin{equation}
\label{subdifmftfull}
\begin{array}{rcl}
\displaystyle\frac{\partial}{\partial t} a(x,t)&=&
~_{0}D_{t}^{1-\gamma }
\left\{K_\gamma\displaystyle\frac{\partial^2}{\partial x^2}
a(x,t)-ka(x,t)b(x,t)\right\} \\
\noalign{\smallskip} \noalign{\smallskip}
\displaystyle\frac{\partial}{\partial t} b(x,t)&=&
~_{0}D_{t}^{1-\gamma }
\left\{K_\gamma\displaystyle\frac{\partial^2}{\partial x^2}
b(x,t)- -ka(x,t)b(x,t)\right\}.
\end{array}
\end{equation}
From the specific reaction term given in Eq.~(\ref{specific}) we
can now obtain the third relation
between the scaling exponents by balancing the terms within the brackets:
\begin{equation}
\mu = 2\theta +1-\gamma.
\label{scaling3}
\end{equation}
Simultaneous solution of Eqs.~(\ref{scaling1}), (\ref{scaling2}), and
(\ref{scaling3}) finally yields
\begin{equation}
\alpha=\frac{\gamma}{6}, \qquad \theta=\frac{\gamma}{3}, \qquad
\mu=1-\frac{\gamma}{3}.
\label{thescales}
\end{equation}

\subsection{Simulated quantities}
\label{model3}
It is useful to list here the quantities that will be compared with
numerical simulations.  Each is characterized by an exponent explicitly
given in terms of $\gamma$. 
The first and second are independent of the choice of
reaction term, but the others are sensitive to this choice.
\begin{enumerate}
\item
The total amount of product $C$ produced as a function of time, given in
Eq.~(\ref{ctot}), is
\begin{equation}
N_C(t)\sim t^{\gamma/2}.
\label{monitor2}
\end{equation}
This scaling is independent of the form of the reaction term.
\item
We measure the width $W_d$ of the depletion zone as the width of the profile
\begin{equation}
U_P(x,t)\equiv 1-|a(x,t)-b(x,t)| .
\label{profile}
\end{equation}
The prediction, which is also independent of the form of
the reaction term, is given in Eq.~(\ref{width}),
\begin{equation}
W_d\sim t^{\gamma/2}.
\label{monitor1}
\end{equation}
\item
We carry out our simulations with an equal initial unit concentration of
$A$ and $B$.
In this case $x_f=0$ for all time.  We monitor the number of $C$ particles
produced at this point of maximum production of $C$, $N_C(x=0,t)$.
Since $R_\gamma(0,t)= dN_C(x=0,t)/dt\sim t^{-\mu}=t^{\gamma/3-1}$, we have
\begin{equation}
N_C(0,t)\sim t^{\gamma/3}.
\label{monitor3}
\end{equation}
This is thus a check on the exponent $\mu$.
\item
The concentration $a(0,t)=b(0,t)$ of each reactant at the center of
the reaction zone is difficult to monitor because it is very small and
therefore subject to large fluctuations.  
Instead, we monitor the integral of this concentration over time,
\begin{equation}
\int_0^t a(0,\tau)d\tau \sim \int_0^t \tau^{-\gamma/3} d\tau \sim
t^{1-\gamma/3}, 
\label{monitor4}
\end{equation}
with a similar result for the other reactant. This then is a check
on the exponent $\theta$.
\item
The width $w(t)$ of the product profile grows, according to the scaling
equation~(\ref{Rgamma}), as $w(t)\sim t^\alpha$.
According to Eq.~(\ref{thescales}) we then have, as a test of $\alpha$, 
\begin{equation}
w(t)\sim t^{\gamma/6}.
\label{monitor5}
\end{equation}
\item
Finally, we monitor the entire profile~(\ref{profile}) as a function of
position and time.
This is a difficult quantity to follow because it involves regions of very
low concentration.  In a way it constitutes a check of the simulation
methodology, as we will see below.
\end{enumerate}

\section{Simulation Details}
\label{sec3}

Monte Carlo simulation methods for reaction-diffusion processes are
ubiquitous.  For a two-dimensional simulation one starts with a
square lattice, and deploys a given number of particles at each site
according to the initial distribution.  The particles then perform a
random walk simulated by the parallel update of the coordinates of all
particles at each time step $t=m\Delta t$, $m=1,2,\cdots$. The entire
lattice is explored at periodic intervals $\Delta t_r$ (which could and
often does coincide with $\Delta t$), and reactions take place at each
site on which there are $A$ and $B$ particles, with probability $kab$.
Here $k\ll 1$ is the reaction rate constant and $a$ and $b$ are 
proportional to the number of particles of type $A$ and $B$ on that
site.  Clearly, $kab$ must (in the appropriate sense, since the
quantity is not dimensionless) be small. 
There are variants of this procedure that are inconsequential for
our analysis (e.g., some excluded volume effects).
A necessary condition to be in the diffusion-controlled 
regime described by the usual reaction-diffusion equations is that the
random walkers on average perform a large number of steps before reacting.  
 
Adjustments that must be made to this procedure in order to describe 
subdiffusion are neither trivial nor straightforward. First and
most importantly, one can not assume that the particles all jump at the
same time.  The distribution of jumping times is now very broad: one can
imagine each particle outfitted with an alarm clock, with a jump to a
randomly selected nearest neighbor taking place when the alarm goes off,
at which time the alarm is reset according to a distribution whose
asymptotic behavior goes as in Eq.~(\ref{waitingtime}). 
Jumping is therefore a renewal process~\cite{cox}.  An example of a
normalized distribution with this behavior is the Pareto law:
\begin{equation}
\psi(t) = \frac{\gamma/\tau}{(1+t/\tau)^{1+\gamma}}.
\label{waitingtimedistrib}
\end{equation}
The particles are labeled, and jumping times are assigned to them
according to this distribution.  These times, from smallest to largest,
must be sorted, and the list must be sorted after each jump or reaction.

Since the particles no longer jump at the same time, a decision must be
made about when they are allowed to react.  There are
at least two alternatives:
(1) A reaction attempt occurs only when a particle first arrives at a site;
(2) Reaction attempts occur at each site at periodic intervals $\Delta
t_r$, and occur with probability $kab$. 
The first alternative does not seem physically reasonable for the
subdiffusive problem since it implies that a pair of $A$ and $B$
particles that remain at a given site and that did not
react upon first encounter will not react no matter how long they remain
at the site, which on average is infinite.  They can only react if
they move apart and then encounter
one another again.  The second alternative, which
we choose for most of our simulations, can be associated with
a number of physical explanations.  On the one hand one can think of
reactions induced or activated periodically by some external agent (a
laser, for example).  More in line with our thinking of subdiffusion as
a way to describe movement in a disordered or glassy or porous medium is 
to think of this as a mesoscopic description.  At a microscopic level
small jumps may occur diffusively, but the motion from one mesoscopic
region to another on a longer time scale is much slower because of
geometric bottlenecks that affect this longer range movement. Our
``sites'' would then correspond to mesoscopic regions in which a walker
can spend a long time moving diffusively from one part of the region
to another.  Reactions can then take place within one of these regions
at regular
time intervals. 

Since the subdiffusive process has a long memory, we must be careful
about the initiation of the process.  In particular, it is not
appropriate to choose the initial jumping times as indicated above
because that would bias the initial condition to one in which all the
particles jumped simultaneously at time $t=0$.  Instead, after this
first selection of times we choose another set of jumping times from the
distribution, and repeat this procedure a large number of times.  The
number of repetitions is usually chosen as the total number of particles
initially in the system.  Only then do we choose $t=0$ by taking the
smallest jumping time as our new origin of time from which the process
is launched.

Finally, it is noted that even in a diffusive process, reaction events
are not really restricted to occur only at periodic time intervals
$\Delta t_r$.  A large literature points to the fact that the continuous
time process underlying such a step process is one in which times are
selected from an exponential distribution~\cite{bedeaux,montroll}
\begin{equation}
P(t)=\kappa e^{-\kappa t},
\label{eq:exponential}
\end{equation}
where $\kappa$ is the reaction rate constant.
We have also tested this procedure in
our subdiffusive system, allowing each pair $(A,B)$ of particles
on one site to react at a
time dictated by such an exponential distribution.  If a particle leaves a
site before a reaction takes place, the reaction ``clock'' of each
particle is reset.  This is also the viewpoint followed by Seki et
al.~\cite{seki1,seki2}.  Note that whereas in the $k, \Delta t_r$
formulation of the reaction events one specifies two parameters, in the
exponentially distributed reaction events the reactions are specified by
the single rate parameter $\kappa$, which identifies both the reaction rate
constant $\kappa$ and the mean time between reaction events
$1/\kappa$.

The parameters used in our simulations are as follows: we place two
particles initially at each site (this corresponds to unit concentration
for each species).  The lattice dimensions are usually $(L_x,L_y)=(20,10)$,
except for $\gamma=3/4$ where we use $L_x=90$, and in some cases
specified later where we use $(L_x,L_y)=(160,20)$.  The
maximum number of particles allowed at a given site is 40.
The rate coefficient is $k=0.05$,  and the time between reaction
events is $\Delta t_r=10$.  For some of the simulations we use 
exponentially distributed reaction events
with $\kappa=10^{-4}$ or $\kappa=10^{-5}$,
which corresponds to a much lower reaction rate. 
The maximum time per run is $t_{max}=1,024,000$.  Results are averaged
over $100$ runs.

\section{Comparisons with Simulations}
\label{sec4}
Here we compare simulations of the six quantities enumerated in
Sec.~\ref{sec:2}
with the theoretical predictions. 
The simulations rapidly become increasingly difficult and time-intensive
with decreasing $\gamma$, and it is
therefore expected that agreement with the theory improves with increasing
$\gamma$.  As we will show, the agreement is on the whole good, especially
for the larger values of $\gamma$. 
We also stress that four of the six comparisons involve results that
decidedly depend on the choice of reaction term. Agreement
would not be obtained with the usual memoryless local law of mass action.

\begin{figure}
\begin{center}
\includegraphics[width=0.6\columnwidth]{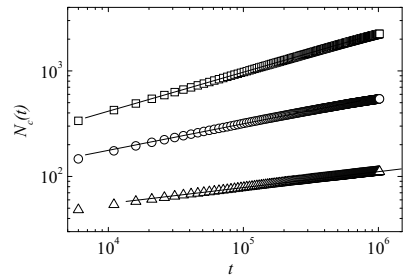}
\end{center}
\caption{Log-log plots of the total number of product particles 
vs time for $\gamma=0.75$ (squares), $\gamma=0.5$ (circles), and
$\gamma=0.25$ (triangles).  The linear fit slopes are $0.37$, $0.25$,
and $0.15$ respectively.
The mean field prediction for the
slope is given in Eq.~(\ref{monitor1}) as $\gamma/2$. \label{fig1}}
\end{figure}

Figure~\ref{fig1} shows our simulation results for the total number of
product particles as a function of time in units of $\tau=1.0$ (used
throughout) for several values of $\gamma$.
This is perhaps the most robust global quantity to be simulated.
The linear fit slope is given for each $\gamma$, and agrees very
well with the theoretical prediction given in Eq.~(\ref{monitor2}) for the
two larger values of $\gamma$.

\begin{figure}
\begin{center}
\includegraphics[width=0.6\columnwidth]{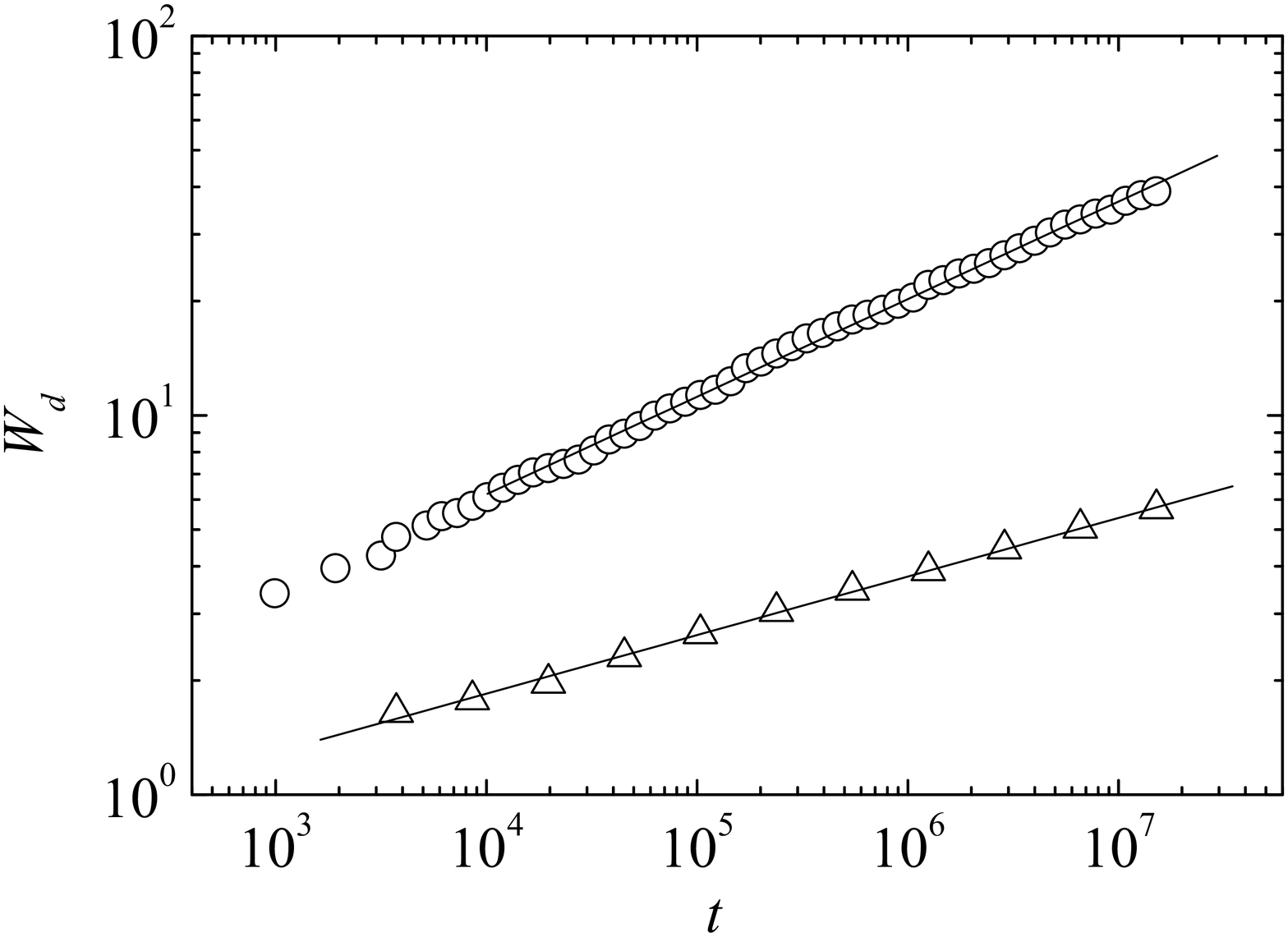}
\end{center}
\caption{Width $W_d$ of the depletion zone vs time 
for $\gamma=0.5$ (circles) and $\gamma=0.25$ (triangles).
The linear fit slopes are $0.257$ and $0.154$ respectively. The mean
field prediction for the
slope is given in Eq.~(\ref{monitor1}) as $\gamma/2$. \label{fig2}}
\end{figure}

Figure~\ref{fig2} shows our simulation results for the width of the
depletion zone as a function for time for two values of $\gamma$.  The
linear fit slope is in good agreement
with the theory as given in Eq.~(\ref{monitor1})
for the larger value of $\gamma$.  Later we discuss some
difficulties, particularly for small values of $\gamma$, in the
accurate simulation of the profile $U_P(x,t)$ whose
width is used to obtain these results.

\begin{figure}
\begin{center}
\includegraphics[width=0.6\columnwidth]{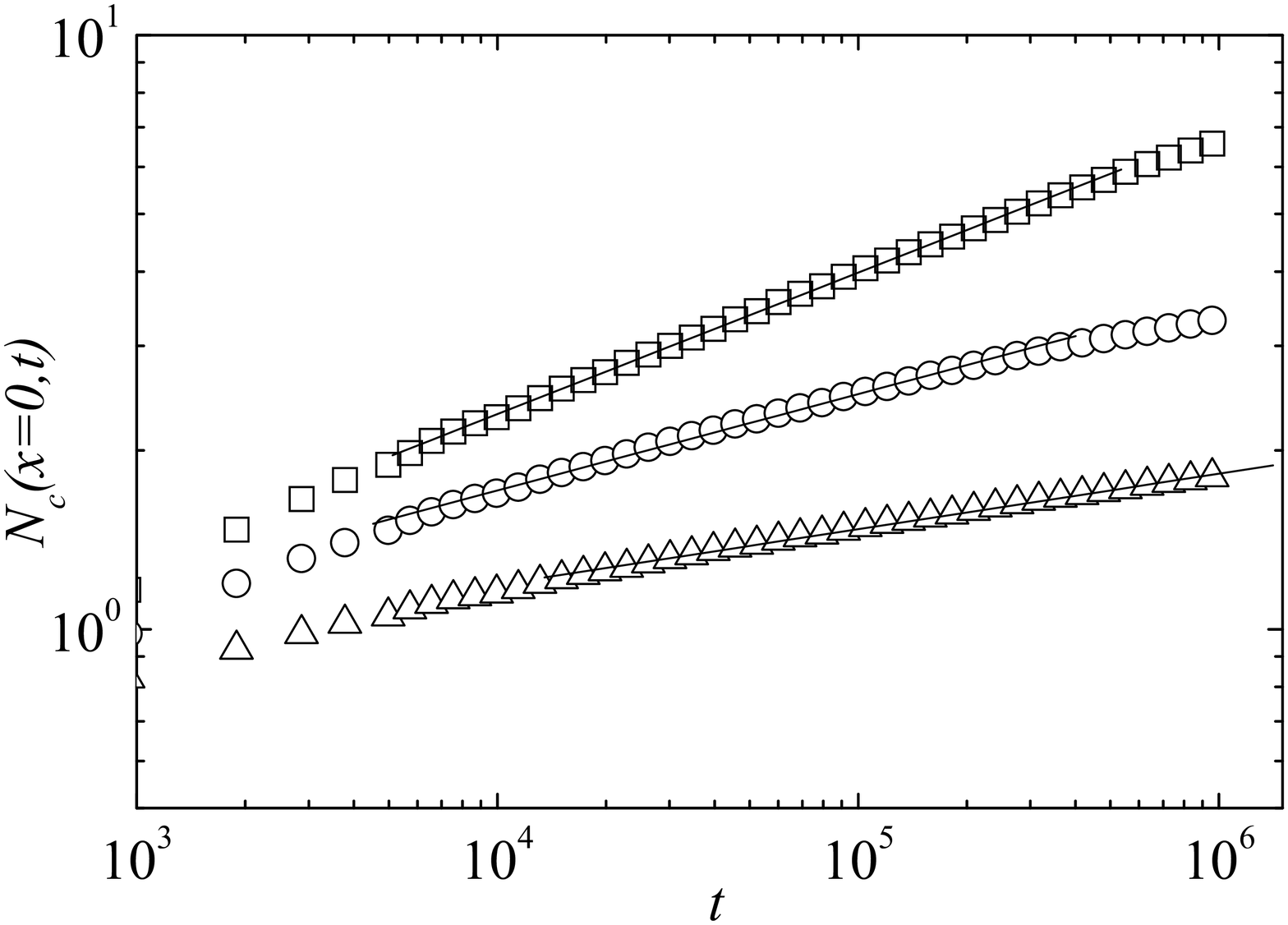}
\end{center}
\caption{Log-log plots of the production profile of product
$C$ at $x=0$ as a function
of time for $\gamma=0.75$ (squares), $\gamma=0.5$ (circles), and
$\gamma=0.25$ (triangles).  The linear fit slopes are $0.24$, $0.162$,
and $0.093$ respectively.
The mean field prediction for the
slope is given in Eq.~(\ref{monitor3}) as $\gamma/3$. \label{fig3}}
\end{figure}

Figure~\ref{fig3} shows our simulation results for the production profile
of the product of the reaction at $x=0$ 
as a function of time, for several values of $\gamma$.
For the larger $\gamma$ the linear fit slope agrees 
well with the theoretical prediction given in Eq.~(\ref{monitor3}).

\begin{figure}
\begin{center}
\vspace*{0.5in}
\includegraphics[width=0.6\columnwidth]{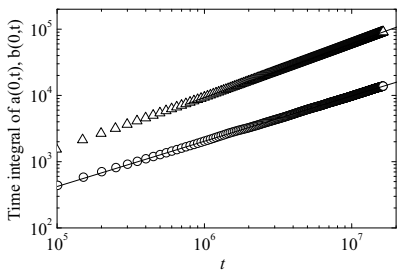}
\end{center}
\caption{Time integrals of reactant concentrations at the center of the
reaction zone.  The mean field prediction for the
slopes is given in Eq.~(\ref{monitor4}) as $1-\gamma/3$.
The steeper curve is for $\gamma=0.25$ and the reaction events
governed by the exponential distribution Eq~(\ref{eq:exponential}) with
$\kappa=10^{-5}$.  The linear fit slope is $0.814$ while mean field
theory yields $0.917$.  The shallower curve is for 
$\gamma=0.5$ and $\kappa=10^{-4}$. The linear fit slope is $0.683$, the 
mean field slope is $0.833$.
\label{fig4}}
\end{figure}

Figure~\ref{fig4} presents our simulation results for the time-integral of
the concentration of a reactant at the center of the reaction zone, to be
compared with the theoretical prediction of Eq.~(\ref{monitor4}). 
While the agreement is not spectacular, the trend is correct.  Also of
interest here is the improved agreement when the reaction rate is
greatly reduced, as expected.
Even more dramatic effects of the reaction rate
are seen below in the context of our discussion of the profile
$U_P(x,t)$.

\begin{figure}
\begin{center}
\includegraphics[width=0.6\columnwidth]{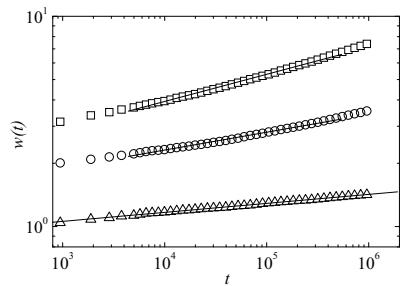}
\end{center}
\caption{Log-log plot of the width of the product profile as a function 
of time for
$\gamma=0.75$ (squares), $\gamma=0.5$ (circles), and
$\gamma=0.25$ (triangles).  The linear fit slopes are $0.129$, $0.084$,
and $0.042$ respectively.  The mean field prediction for the
slope is given in Eq.~(\ref{monitor5}) as $\gamma/6$. \label{fig5}}
\end{figure}

\vspace*{0.5in}
Figure~\ref{fig5} contains our simulation results for the width of the
product profile, which should be compared with the prediction of 
Eq.~(\ref{monitor5}). The agreement is very good for all the values of
$\gamma$.

\begin{figure}
\begin{center}
\includegraphics[width=0.6\columnwidth]{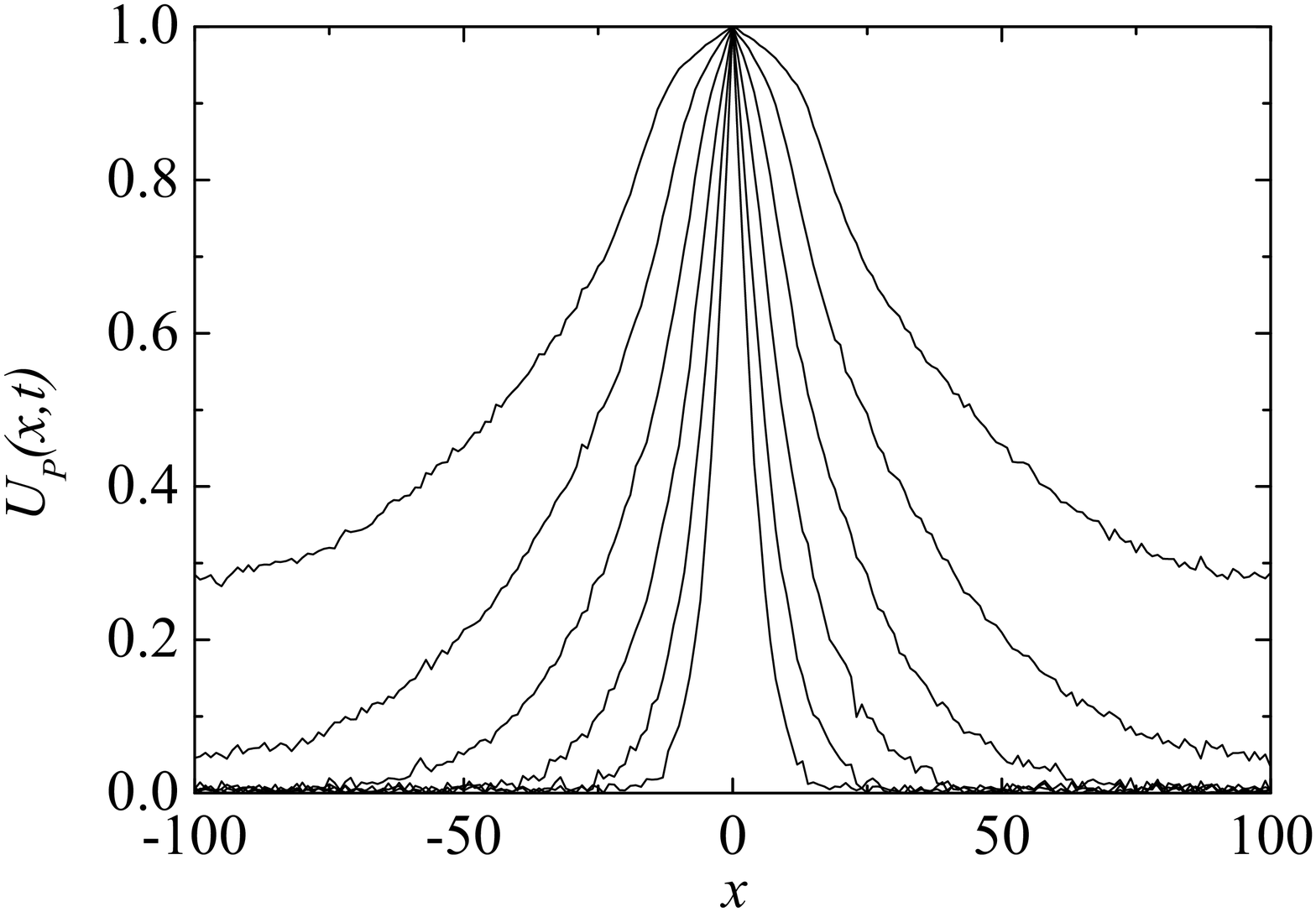}
\end{center}
\caption{
Simulation results for the profile $U_P(x,t)$ for
$\gamma=0.75$ and $t=946$, $3777$, $15073$, $60149$, $240025$, and
$957828$. The width of the profiles increase with time, as seen in
Fig.~\ref{fig5}.
The simulation was carried out on a lattice of size $(160,20)$ and
averaged over $102$ runs. 
Notice the apparent
evolution from a characteristic sharp-pointed profile for short
times to a vaulted profile at longer times. For a discussion of 
this anomaly, and for the values of other parameters, see
text.
}
\label{fig6}
\end{figure}

\begin{figure}
\begin{center}
\includegraphics[width=0.8\columnwidth]{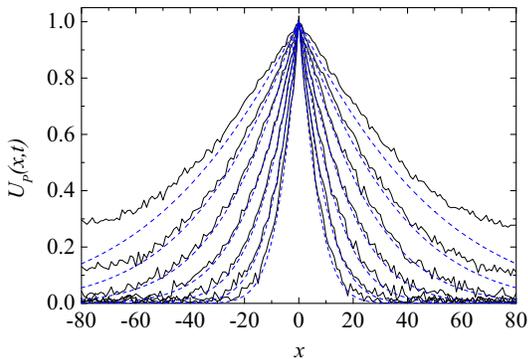}
\end{center}
\caption{
Simulation (jagged lines) and theory (dashed lines) for the
profile $U_P(x,t)$ for
$\gamma=0.75$ and $t=3738$, $8577$, $19682$, $45165$, $103638$,
$237815$, and $545708$.  
Notice the absence of the rounding anomalies in the
profile for small $x$. The reaction rate is much smaller in this case,
see text. 
The simulation was carried out on a lattice of size $(160,20)$ and
averaged over $102$ runs. 
The times are approximately equivalent
to those used in Fig.~\ref{fig6} if the total number of $C$  
particles in the system is used as a measure of time.
}
\label{fig7}
\end{figure}

Finally, in Figs.~\ref{fig6} and \ref{fig7} we present perhaps the
most difficult quantity
to capture accurately, namely, the profile $U_p(x,t)$ defined in
Eq.~(\ref{profile}).  For these simulations we use a lattice of size
$(L_x,L_y)=(160,20)$.
As pointed out earlier, the difficulty arises
from the fact that it involves regions of very low concentration.  It is
instructive to illustrate the difficulty, and that is why we have
included Fig.~\ref{fig6}.  The simulation
profiles shown at the different times are
actually time averaged over a small time interval around the times
shown. The noteworthy feature is the evolution of the sharp-pointed
profile near the origin at short times to a more rounded shape at
longer times.  The mean field theory presented in this paper does not
produce this rounding, so that it seemed at first that the theory
and simulations
differed in some profound way.  However, the simulations in
Fig.~\ref{fig6} were carried out with the rate coefficient $k=0.05$ with
reactions occurring periodically at time interval $\Delta t_r=10$, a
reaction rate that turns out to be too high for comparison with our
theory.  In Fig.~\ref{fig7} we show both the simulation results
(jagged curves) and those of our theory (dashed curves),
now with exponentially distributed reaction
events according to Eq.~(\ref{eq:exponential}), with $\kappa=10^{-5}$.
The profile near $x=0$ remains pointed for all the times shown,
as predicted by the theory.
The quantitative disagreements for long times ($t=237185$ and
especially $t=545708$) are due to finite size effects.
Boundary effects are negligible only as long as $U_p(x=-L_x/2,t)=
U_P(x=L_x/2,t)=0$, and for the long times we find that this is not the
case.  In our other simulation results we have not included such results
in our averages, but have left them in this figure simply to stress
the finite
size effects one must be aware of in calculating the behavior of
quantities in the depletion zone when the zone extends all the way to
the boundaries of the system.  To provide accurate results for such long
times it is necessary to run simulations on larger systems.

\section{Closing Comments}
\label{sec5}
In this paper we have proposed a set of continuum fractional diffusion
equations to describe the behavior of a reaction front in the $A+B \to C$ 
\emph{reaction-subdiffusion} problem.  Subdiffusion may be
appropriate to describe the way reactants move in complex
(glassy, disordered, highly constrained) geometries, and we were
interested in exploring how this constraint on the motion would affect
the evolution of a reaction front.  Because we are working with a set of
mean field continuum equations, our results are only valid above the
critical dimension $d=2$.  

The subdiffusive motion is modeled via the usual fractional equation
that contains the Riemann-Liouville operator, Eq.~(\ref{LR}).  This
choice has a long history, and its virtues and shortcomings are clearly
understood.  Less clear has been the selection of the local reaction
term, and the question of the way in which the memory in the
Riemann-Liouville operator does (or does not) affect the way in
which the reaction is modeled.  While the literature on this subject has
presented a number of viewpoints, we argued, in agreement
with~\cite{seki1,seki2}, that the reaction term should
also be modified from its usual simple instantaneous product
form, at least for small reaction rate constants. 
Our reaction-subdiffusion model is thus given by
Eq.~(\ref{subdifmftfull}). 

Following the approach of G\'alfi and R\'acz~\cite{galfi} for the
evolution of a front in the reaction-diffusion problem, we assumed 
scaling solutions for the various quantities that can be calculated from
the model. Some of these quantities depend explicitly on the form
chosen for the reaction term while others do not.  We compared the
resulting exponents with those obtained from numerical simulations.  We
found very good agreement between the theory and simulations for
the exponents $\mu$ and $\alpha$ that characterize the
reaction term, Eq.~(\ref{Rgamma}).  In particular, in terms of the
power $\gamma<1$ that characterizes the subdiffusive process we found that
the theoretical values $\mu=1-\gamma/3$ and $\alpha=\gamma/6$ are
recovered in the simulations with greater fidelity for larger $\gamma$. 
The exponent
$\theta$, governing the time decay of the reactant concentrations
as in Eq.~(\ref{abscaling}),
is, theoretically, given by $\gamma/3$. Simulation results give
values which correctly follow this trend, but the agreement is
not quantitative. However, we have to remember that the
quantity $a(0,t)$ is local and, consequently, it is more difficult
to achieve good statistical averages with the small systems and
number of particles considered in the simulations.  Perhaps the most
challenging quantity to capture is the profile $U_P(x,t)$. The theory
predicts a cusp at $x=0$ which we were able to capture by our
simulations when the reaction rate constant is sufficiently
small.  The quantitative agreement between the theory and the
simulations for this profile was ultimately limited by our finite system
size.  We note that our results are a good example of
what is sometimes referred to as subordination in that the
subdiffusive scaling behavior
can be deduced from the corresponding diffusive behavior with the
substitution $t\to t^\gamma$~\cite{subordination}.

This work can clearly be pursued along a number of directions. 
Among them is the description of this same reaction-subdiffusion problem
with the usual uniform initial condition for the species, to investigate
what sorts of segregation patterns might evolve on the way to
extinction, or on the way to equilibration if the reaction is reversible.
Another is the study of the fluctuations that must be added to the model
in order to describe this process in a one-dimensional system where the
mean field description is no longer appropriate, and the
possible logarithmic correction in two dimensions that may explain some of
our small-$\gamma$ discrepancies.  A third is the effect of different
subdiffusion coefficients for the species $A$ and $B$, and even of
different exponents $\gamma_A$ and $\gamma_B$.  In this latter case 
subordination would necessarily be more complicated if valid
at all.  Work along these directions is in progress~\cite{newones}.

\begin{acknowledgments}

This work was partially supported by the Ministerio de Ciencia y
Tecnolog\'{\i}a (Spain) through Grant No. BFM2001-0718
and by the Engineering Research Program of
the Office of Basic Energy Sciences at the U. S. Department of Energy
under Grant No. DE-FG03-86ER13606.
\end{acknowledgments}

\end{document}